\documentclass[aps,epsfig,rotate,showpacs,preprint]{revtex4}
\usepackage{epsfig}
\usepackage{graphicx}
\usepackage{amsmath}

\newcommand{\beq}{\begin{eqnarray}}
\newcommand{\eeq}{\end{eqnarray}}

\def\ak{\hat{a_{{\bf k}}}}
\def\akh{\hat{a_{{\bf k}}}^{\dagger}}

\def\Smatrix{S_{ABCD}}
\def\psia{|\phi^{A}_{E_{A}, n_{A}}\rangle}
\def\psib{|\phi^{B}_{E_{B}, n_{B}}\rangle}
\def\psic{|\phi^{C}_{E_{C}, n_{C}}\rangle}
\def\psid{|\phi^{D}_{E_{D}, n_{D}}\rangle}

\begin{document}

\title{Indistinguishability and Interference in the Coherent Control of Atomic and Molecular Processes}
\author{Jiangbin Gong}
\affiliation{Department of Physics and Centre for Computational Science and Engineering,
National University of Singapore, 117542, Republic of Singapore}
\affiliation{NUS Graduate School for Integrative Sciences and Engineering, 117597,
Republic of Singapore}
\author{Paul Brumer}
\affiliation{Chemical Physics Theory Group, Department of Chemistry \\ and
Center for Quantum Information and Quantum Control,\\
University of Toronto, Toronto, M5S 3H6, Canada}
\date{\today}
\begin{abstract}
The subtle and fundamental issue of indistinguishability and
interference between independent pathways to the same target state
is examined in the context
of coherent control of atomic and molecular processes, with emphasis
placed on possible ``which-way" information due to
quantum entanglement established in the quantum dynamics.
Because quantum interference between independent pathways
to the same target state occurs only when the
independent pathways are indistinguishable, it is first shown
that creating useful
coherence between nondegenerate states of a molecule
for subsequent quantum interference manipulation
cannot be achieved by collisions between atoms or molecules that are prepared in momentum and energy eigenstates.
Coherence can, however,  be transferred from light fields to
atoms or molecules.
Using a particular coherent control scenario, it is shown that
this coherence transfer
and the subsequent coherent phase control
can be readily realized
by the most classical states of light, i.e.,
coherent states of light. It is further
demonstrated that
quantum states
of light may suppress the extent of phase-sensitive coherent control by leaking
out
some which-way information while
``incoherent interference control" scenarios proposed in the literature
have automatically ensured the indistinguishability of
multiple excitation pathways.
The possibility of quantum coherence in photodissociation
product states is also understood in terms of the disentanglement
between photodissociation fragments.
Results offer deeper insights into quantum coherence generation in atomic and
molecular processes.
\end{abstract}
\maketitle
\section{Introduction}
\label{s4-1}

Interference constitutes one of the most intriguing aspects of
quantum mechanics \cite{silverman,greenstein}. Specifically, when a
target state is accessible by two or more independent pathways, the
corresponding probability amplitudes must be added, rather than the
corresponding probabilities added, in order to obtain the overall
result. The realization that the resulting constructive or
destructive interference terms can be altered via experimentally
controllable parameters motivated the development of the field of
``coherent control'' \cite{brumerbook,rice,rabitz}. Considerable
theoretical and experimental work has been done in this area,
demonstrating our ability to actively control  atomic and molecular
processes by manipulating quantum interference.

Despite the fact that quantum interference allows for active
control, the mystery it contains, however,  admittedly remains
\cite{feynman,jammer,qcbook}. This becomes particularly  clear if we
consider the role of Young's double-slit interference experiment for
atoms or electrons, introduced during the early days of  quantum
mechanics. Traditional understandings of the double-slit experiment
are based largely on Bohr's complementarity, i.e. wave-particle
duality.  That is, if we can measure the recoiling slits and
successfully determine which slit a particular particle goes
through, then the interference will be necessarily washed out due to
the position uncertainty introduced by measurement. As a result, the
wave-like property and the particle-like property cannot be observed
simultaneously. Modern quantum optical tests further support Bohr's
claim \cite{scully}.  For example, in the double-slit experiments
using atomic beams,  micromaser detectors can determine which path
the particle has followed without affecting the atomic spatial
wavefunction by use of the atomic internal degrees of freedom. Doing
so, however, causes loss of interference as a consequence of the
available which-way information contained in the entanglement
between the particles and the measuring apparatus
\cite{schlosshauer}. Other one-photon and two-photon interference
experiments \cite{mandel} further  demonstrated that quantum
interference should be understood as the  physical manifestation of
the intrinsic indistinguishability of multiple independent pathways,
and that once there is some way, even in principle, of
distinguishing between independent quantum routes to the same target
state, the corresponding probabilities should be added and quantum
interference no longer exists.

There exist some useful quantitative treatments
that relate quantum interference to the
indistinguishability between multiple pathways
\cite{zurek41,uffink,englert,mandel2}.
Because they are not the focus of this paper,
here we briefly introduce only
one of them that will be occasionally used below.
Let $|\psi_{1}\rangle$
and $|\psi_{2}\rangle$ denote the final states associated with
two pathways.
A measure for
distinguishing between these two states can be described by a complete
set of commuting orthogonal projection operators $\{ P_{n}\equiv
|\xi_{n}\rangle\langle \xi_{n}| \}$.
The degree of {\em indistinguishability}
can then be defined as
\begin{eqnarray}
U_{P_{n}}(\psi_{1},
\psi_{2})=
\sum_{n}\sqrt{\frac{\langle\psi_{1}|P_{n}|\psi_{1}
\rangle\langle\psi_{2}|P_{n}|\psi_{2}\rangle}{\langle\psi_{1}|\psi_{1}
\rangle\langle\psi_{2}|\psi_{2}\rangle}},
\label{defineu}
\end{eqnarray}
where $U_{P_{n}}(\psi_{1}, \psi_{2})=1$ represents
maximal indistinguishability (e.g., when $|\psi_{1}\rangle$
differs from $|\psi_{2}\rangle$ by a phase factor only),
and
$U_{P_{n}}(\psi_{1}, \psi_{2})=0$ represents
perfect distinguishability.
On the other hand, the degree of
{\em interference power}
of $|\psi_{1}\rangle$ and $|\psi_{2}\rangle$, for a complete
set of orthogonal projection operators $\{ P'_{l}\equiv
|\eta_{l}\rangle\langle\eta_{l}|\} $, is defined as
\begin{eqnarray}
I_{P'_{l}}(\psi_{1}, \psi_{2})=\sum_{l}
\frac{|\langle\psi_{1}|P'_{l}|\psi_{2}\rangle|}{
\sqrt{\langle\psi_{1}|\psi_{1}\rangle
\langle\psi_{2}
|\psi_{2}\rangle}},
\label{defineI}
\end{eqnarray}
It has been proved that if $\{P_{n}\}$ and $\{P'_{l}\}$ commute,
then the degree of interference power is always no larger than that
of indistinguishability \cite{uffink}, i.e.,
\begin{eqnarray}
U_{P_{n}}(\psi_{1}, \psi_{2})\geq I_{P'_{l}}(\psi_{1},\psi_{2}).
\label{uffink-eq}
\end{eqnarray}

Equation (\ref{uffink-eq}) implies that entangling the initial
system wavefunctions $|1\rangle$, $|2\rangle$ with measurement
apparatus wavefunctions  $|\xi_{1}\rangle$, $|\xi_{2}\rangle$ should
unavoidably affect quantum interference contriburions
\cite{steinberg}. To see this more clearly let us assume that the
decoupled time evolution of the system and the measurement apparatus
are described by the unitary operators $U_{S}$ and $U_{M}$,
respectively. The corresponding final states for the system and the
measurement apparatus are given by
$|\psi_{1}\rangle=U_{S}|1\rangle\otimes U_{M}|\xi_{1}\rangle $  and
$|\psi_{2}\rangle=U_{S}|2\rangle\otimes U_{M}|\xi_{2}\rangle $.
Then, the degree of interference power [Eq. (\ref{defineI})] for any
measurement $P'_{l}$ done on the system only is proportional to
$|\langle \xi_{1}|U_{M}^{\dagger}U_{M}|\xi_{2}\rangle|=|\langle
\xi_{1}|\xi_{2}\rangle|$. Hence,  when $|1\rangle$ and $|2\rangle$
are entangled with orthogonal states $|\xi_{1}\rangle$ and
$|\xi_{2}\rangle$, there  will be no interference because $\langle
\xi_{1}|\xi_{2}\rangle=0$. Indeed, the degree of
indistinguishability [Eq. (\ref{defineu})] between
$|\psi_{1}\rangle$ and $|\psi_{2}\rangle$  is zero if we choose
$P_{1}=|\xi_{1}\rangle\langle \xi_{1}|$ and
$P_{2}=|\xi_{2}\rangle\langle \xi_{2}|$. That is, the measurement of
$P_{1}$ and $P_{2}$ serves to distinguish between $|1\rangle$ and
$|2\rangle$, and thus to distinguish between the two pathways
$|1\rangle \rightarrow |\eta_{l}\rangle$ and $|2\rangle\rightarrow
|\eta_{l}\rangle$. On the other hand, if the states
$|\xi_{1}\rangle$ and $|\xi_{2}\rangle$ are similar (i.e., their
overlap is considerable),  then the entanglement between the system
and the measuring apparatus is weak enough to allow for quantum
interference (i.e., the measurement of $P_{1}$ and $P_{2}$ will not
provide much which-way information about the system and interference
will not be lost) \cite{schlosshauer}.

Qualitative aspects of the
relationship between interference and which-way information are
crucial to the entire discussion below.

In the context of coherent control,  our appreciation of the
usefulness of interference far exceeds our understanding of the
fundamental essence of interference. In particular, in laser control
of atomic and molecular processes, a laser pulse is usually employed
to prepare a superposition state for the subsequent generation of
multiple coherent pathways, analogous to the role of a beam splitter
in matter-wave interference experiments. Fundamentally interesting
questions then arise. For example, why do laser fields provide an
important means of creating molecular coherence for subsequent
quantum interference control? Is there any which-way information
extractable from the light field after the light-matter interaction
is over? Are the independent pathways indeed largely
indistinguishable so as to ensure interference?  Further, with
regard to indistinguishability between multiple excitation pathways,
how do we understand  the well-known fact that, on one hand
traditional interference control scenarios are extremely
phase-sensitive,  and on the other hand there are also some
phase-insensitive interference control scenarios
\cite{brumerbook,rice}?

Questions such as these have motivated us to examine the
issue of indistinguishability
and interference in the specific context of the coherent
control of atomic and molecular processes.  Our goal is to
obtain deeper
insights into the nature of coherence and interference in coherent control and to establish
some foundational concepts in this area.

This paper is organized as follows.
Section \ref{s4-2} presents considerations
on the possibility of creating useful quantum coherence
via atomic or molecular collisions.  We show that,
due to indistinguishability requirements,
collisions between atoms or molecules prepared in momentum and energy eigenstates cannot, in almost all instances,
create useful quantum coherence for subsequent control.
Thus, as the first step in most quantum control scenarios, creating a
superposition state comprising nondegenerate states is done
by coherent laser irradiation.  In Sec.
\ref{s4-3}, we (1) compare a fully quantized theory of
a two-pulse coherent control scenario with a classical-field treatment,
in order to understand conditions under
which laser-molecule interaction can create molecular
coherence without leaking out which-way information; and (2) analyze
the differences between phase sensitive and phase insensitive
control scenarios in terms of the nature of
indistinguishability of multiple excitation pathways.
The creation of quantum coherence
in photodissociation processes is discussed in Sec. \ref{s4-4}.
We briefly summarize and conclude this work
in Sec. \ref{s4-5}.

\section{On Quantum Coherence Creation By Atomic and Molecular Collisions}
\label{s4-2}

In several important coherent control scenarios \cite{brumerbook,rice},
the first step is to create a superposition state that results in
multiple coherent pathways to the same  target state. Here we
examine the possibility of preparing useful quantum superposition
states via atomic and molecular collisions. To see that this is a
rather fundamental issue, consider first a typical molecular
crossed-beam experiment (to be specific we focus below on
bimolecular collisions, but these considerations apply to atomic
collisions as well).  Because the interaction time $\delta t$ in the
crossed-beam experiment is short, in effect each beam is subject to
a ``pulsed" interaction due to the other beam.  It would then appear
that such a molecular process is analogous to pulsed laser-molecule
interaction, and may be able to generate useful quantum
superposition states with the characteristic energy coherence width
given by $\hbar/\delta t$.   However, this turns out to be incorrect
even when the total system is perfectly isolated from the
environment. As will become evident,  this is because different
energy or momentum components of the quantum state of one particle
of interest are distinguishable by measuring the quantum state of
other particles.  This feature is directly related to
EPR (Einstein-Podolsky-Rosen) arguments \cite{EPR}.

Consider the bimolecular collision,
$A + B$ $\rightarrow$ $C + D$,
where $A,B,C,D$ are, in general, molecules of mass $m_A$, $m_B$, $m_C$
and $m_D$. Their free internal Hamiltonians are $H_{A}^{0}$, $H_{B}^{0}$,
$H_{C}^{0}$ and $H_{D}^{0}$.
Here $C$ and $D$ can be identical
to $A$ and $B$ (nonreactive scattering) or can differ from $A$ and $B$
(reactive scattering).
For a molecule denoted by $X$, the momentum
eigenket (in the laboratory frame) is denoted by
$|{\bf K}_{X}\rangle$
with the wave vector ${\bf K}_{X}$, and the internal eigenstate is represented by
$|\phi^{X}_{E_{X},n_{X}}\rangle$, where $E_{X}$ is the internal
energy of particle $X$ and $n_{X}$ is
a set of good quantum numbers accounting for degeneracies.
The eigenkets associated with the relative motion and the
center of mass of motion of two molecules $X$ and $Y$
are denoted by
$|{\bf k}_{XY}^{E_{X},E_{Y}}\rangle$ and $|{\bf K}_{XY}\rangle$.
Now suppose that the preparation of a
superposition state of the molecule $C$ is of particular interest. Then
other channels not including $C$ can be neglected and
all possible states
$\psic\otimes\psid\otimes|{\bf k}_{CD}^{E_{C},E_{D}}\rangle\otimes|{\bf K}_{CD}\rangle$
form a complete basis set of the Hilbert space of interest.

We assume that initially molecules $A$ and $B$ are in pure internal states
$\psia$ and $\psib$. Under normal circumstances
their translational states can be taken as  momentum eigenstates
if they are created in a molecular beam apparatus \cite{note42}. In addition, since one of our goals is to see
if coherence between different momentum or energy eigenstates can be created by a collision process, we naturally assume that
the initial states are prepared as momentum and energy eigenstates.
Thus, before the collision the quantum state is
\beq
|\Psi_{0}\rangle &= & \psia\otimes\psib\otimes|{\bf k}_{AB}^{E_{A},E_{B}}\rangle\otimes|{\bf
K}_{AB}\rangle,
\eeq
where
${\bf k}_{AB}^{E_{A},E_{B}}=(m_{B}{\bf k}_{A}-m_{A}{\bf k}_{B})/(m_{A}+m_{B})$, ${\bf K}_{AB}={\bf K}_{A}+{\bf K}_{B}$.
After the collision this initial state evolves to
$|\Psi_{CD}\rangle$ for the product channel $C+D$.

Of interest is whether a bimolecular collision of this kind produces
any useful coherence
in molecule $C$. By useful coherence we mean coherence in $C$ that could
be used  for a subsequent coherent control scenario, without reference to
molecule $D$.
To address this issue the quantum state composition of $|\Psi_{CD}\rangle$
needs to be examined. Specifically,
$|\Psi_{CD}\rangle$ can be written as
\begin{eqnarray}
  |\Psi_{CD}\rangle &  = &
 \int d\Omega\sum_{E_{C}, n_{C}}\sum_{E_{D}, n_{D}}\Smatrix(E_{C},
n_{C}; E_{D}, n_{D}|\Omega) \nonumber \\
& & \times\  \psic\otimes|{\bf K}_{C}
\rangle\otimes\psid\otimes|{\bf K}_{D}\rangle,
\label{psicomposition}
\end{eqnarray}
where $\Smatrix(E_{C}, n_{C}; E_{D}, n_{D}|\Omega)$ denotes an
on-shell scattering matrix, the direction of ${\bf
k}_{CD}^{E_{C},E_D}$ is given by $\Omega$, and the magnitude of
${\bf k}_{CD}^{E_{C},E_D}$ is uniquely determined by momentum and
energy conservation. As is seen clearly from  Eq.
(\ref{psicomposition}), a collision between $A$ and $B$
can readily produce a superposition state consisting of different
translational and/or internal quantum states, but such a
superposition state is in the entire product space of $C$ and $D$,
i.e., both the translational and the internal states of $C$ and $D$
are entangled. Such entanglement between molecules $C$ and $D$ can,
in principle, provide which-state information about the molecule
$C$. Indeed, if we have obtained the values of ${\bf K}_{D}$ and
$E_{D}$ by measuring $D$ only,  then  from momentum and energy
conservation of the collision process we can infer ${\bf K}_{C}$ and
$E_{C}$. Hence, a natural set of projection operators for
distinguishing between different $C$ states, and thus distinguishing
between different pathways associated with these states,  are given
by
\begin{eqnarray}
P_{{\bf K}_{D},E_{D},n_{D}}=|{\bf K}_{D}\rangle|\phi^{D}_{E_{D},n_{D}}\rangle
\langle\phi^{D}_{E_{D},n_{D}}|\langle {\bf K}_{D}|.
\label{dis-set}
\end{eqnarray}
As one can now readily check,
the corresponding degree of indistinguishability between
the $C$ states with different $E_{C}$ or ${\bf K}_{C}$ is zero.
The interference power of the $C$ states is hence zero, i.e.,
the $C$ states with different $E_{C}$ or ${\bf K}_{C}$ cannot be used
for coherent control in a second process that does not involve molecule
$D$.

Clearly, if two particles ($C$ and $D$) have once interacted with
each other, they will never be separable. It is this
nonseparability, the essence of the EPR argument \cite{EPR}, that
makes which-state information about one molecule available. Thus,
the availability of the which-state information about $C$ rules out
bimolecular collisions as a rational means of creating
indistinguishable multiple pathways (associated with $C$) to the
same target state. To our knowledge, this seemingly straightforward
result is not well appreciated even in considerations of atom and
molecule interferometry \cite{gongprl}.

To further understand the relationship between
interference and indistinguishability, and to
examine whether or not there are useful coherence effects
between degenerate internal states of $C$,
we alternatively use below a density matrix approach.
The density operator $\hat{\rho}_{CD}$ in the $C+D$ arrangement
after the preparation  collision is given  by
\begin{eqnarray}
 \hat{\rho}_{CD}&\equiv&
|\Psi_{CD}\rangle\langle\Psi_{CD}|
=\int\int d\Omega d\Omega'
\sum_{E_{C}, n_{C},E_{D}, n_{D}}\ \sum_{E'_{C}, n'_{C}, E'_{D}, n'_{D}}
\nonumber \\
 & & \Smatrix(E_{C},n_{C};E_{D},n_{D}|\Omega)
 \Smatrix^{*}(E'_{C},n'_{C};E'_{D},n'_{D}|\Omega') \nonumber \\
& & \times\ \psic\otimes|{\bf K}^{E_{C}, E_{D}}_{C}\rangle\otimes\psid
\otimes|{\bf K}_{D}^{E_{C}, E_{D}}
\rangle \nonumber \\
& &\ \ \ \langle {\bf K}_{D}^{E_{C}', E_{D}'}|\otimes\langle\phi^{D}_{E_{D}',n_{D}'}|\otimes
\langle {\bf K}^{E_{C}', E_{D}'}_{C}|\otimes\langle
\phi^{C}_{E_{C}',n_{C}'}|.
\label{rhoprep}
\end{eqnarray}
Consider then a second collision between molecule $C$ and another
object $F$ that may be an atom, a molecule or a photon. The total
initial density operator $\hat{\rho}_{i}$ for the second collision
with regard to the entire system is a product of the density
operator prepared in the first collision and the density operator
$\hat{\rho}_{F}$ of $F$, i.e., $\hat{\rho}_{i}\equiv
\hat{\rho}_{CD}\otimes \hat{\rho}_{F}$. If the second scattering
process is described by the scattering operator ${\bf \hat{S'}}$,
then the total density operator after the second process is given by
${\bf \hat{S'}}^{\dagger}\hat{\rho}_{CD}{\bf \hat{S'}}$. Because the
second scattering process is assumed to be independent of molecule
$D$, all operators involving only $D$ commute with the scattering
operator $ {\bf \hat{S'}}$.  Furthermore, the projector associated
with a target state, denoted $|\phi\rangle\langle\phi|$, is also
assumed to be independent of molecule $D$ and therefore commutes
with operators involving $D$ only. After first tracing over those
degrees of freedom associated with $D$, we then obtain that the
probability $P_{t}$ of reaching the target state is given by
\begin{eqnarray}
P_{t} &=&\int d\Omega\sum_{E_{C},E_{D},
n_{D}}\sum_{n_{C}, n_{C}'}
\Smatrix(E_{C},n_{C};E_{D},n_{D}|\Omega)
 \Smatrix^{*}(E_{C},n_{C}';E_{D},n_{D}|\Omega) \nonumber \\
& & \times\ {\hat{T}}_{ND}\left(
{\bf \hat{S'}}^{\dagger} \psic\otimes|{\bf K}^{E_{C}, E_{D}}_{C}\rangle\langle
{\bf K}^{E_{C}, E_{D}}_{C}|\otimes\langle\phi^{C}_{E_{C},n_{C}'}|\otimes \hat{\rho}_{E}{\bf
\hat{S'}}|\phi\rangle\langle\phi|\right)\nonumber \\
& \equiv &
\int d\Omega\sum_{E_{C},E_{D},
n_{D}}\sum_{n_{C}, n_{C}'}
 \Smatrix(E_{C},n_{C};E_{D},n_{D}|\Omega)
 \Smatrix^{*}(E_{C},n'_{C};E_{D},n_{D}|\Omega)
T^{n_{C}, n_{C}'}_{E_{C}, E_{D},
n_{D}}(\Omega), \nonumber
\\
\label{result}
\end{eqnarray}
where $\hat{T}_{ND}$ denotes the trace over all degrees of freedom
excluding molecule $D$. Note also that Eq. (\ref{result}) also
defines $T^{n_{C}, n_{C}'}_{E_{C}, E_{D}, n_{D}}(\Omega)$.

Note first that,  $T^{n_{C}, n_{C}'}_{E_{C}, E_{D}, n_{D}}(\Omega)$
only contains terms
diagonal in the translational motion of $C$. Hence, useful translational
coherence of $C$ is not generated.
Likewise, the probability of reaching the target state is diagonal in representation of the
nondegenerate internal states of $C$.
Thus, the product of the $C+F$ collision does not see the
``quantum coherence'' between nondegenerate states of $C$.
Clearly, here one can  expect no interference whatsoever between
multiple pathways $(E_{A},E_{B},n_{A},n_{B}, {\bf
k}_{AB}^{E_{A},E_{B}})$$
\rightarrow [E_{C}(E_{C}'),n_{C}, {\bf K}_{C}^{E_{C},E_{D}}({\bf
K}_{C}^{E_{C}',E_{D}'})]$
$\rightarrow $ target states.

These results are consistent with the previous discussion  based on
the indistinguishability requirement.  Interestingly,
in the derivation here the degrees of freedom associated with $D$
are traced over since $D$ is not of interest.
As such, there are no useful quantum coherences between
the $C$ states with different momentum or internal energy,
even though we do not make any
measurement on $D$ to distinguish between the $C$ states.
This observation provides
greater insight into the issue. That is, it makes clear that
what is fundamentally important is
not that we actually distinguish between different $C$ states by
use of the projectors $P_{{\bf
K}_{D},E_{D},n_{D}}$ defined in Eq. (\ref{dis-set}), but rather that
there exists the {\em possibility}
of distinguishing between different $C$ states.  This is an
example that, as far as indistinguishability between independent pathways
is concerned, what
counts is what you can do, not what you actually do \cite{greenberger,greenstein}.

The which-state information about $C$ is seen to arise as a direct consequence
of the conservation laws of total momentum and  total energy of $C$
and $D$. To ``erase'' the which-state information afforded by the
quantum entanglement (and hence to allow for the possibility of interference),
we need to introduce uncertainty in knowing
the state of $C$ from measuring $D$, i.e., it is necessary to
introduce some uncertainties in momentum or energy into the system.
For example,  (1) momentum uncertainties may be easily introduced
to molecular systems by collisions between molecules and macroscopic
objects that in general have considerable momentum variances as
compared with microscopic particles \cite{gongprl}; and (2) as shown
below, energy uncertainties can be easily introduced through laser
excitation. It is for the second reason that laser technologies are so important
to coherent control. Bimolecular
collisions themselves are seen not a useful means for producing interesting
quantum coherence effects for subsequent manipulation of
interference effects.

The above straightforward density matrix formalism also allows us to
examine if bimolecular collisions may create quantum coherence
between {\em degenerate} molecular states of $C$. Interestingly, in
Eq. (\ref{result}), one sees that for each specified set of $\Omega,
E_{C}, E_{D}, n_{D}$, the reaction probability contains the diagonal
terms $ T^{n_{C}, n_{C}}_{E_{C}, E_{D}, n_{D}}(\Omega)$, and the
cross terms $T^{n_{C}, n_{C}'}_{E_{C}, E_{D}, n_{D}}(\Omega)$
$(n_{C}\neq n_{C}')$. The cross terms describe interference effects
between the pathways $(E_{A},E_{B},n_{A},n_{B}, {\bf
k}_{AB}^{E_{A},E_{B}})$ $\rightarrow$ $ [E_{C},n_{C}(n_{C}'), {\bf
K}_{C}^{E_{C},E_{D}}, n_{D}, {\bf K}_{D}^{E_{C},E_{D}}]$
$\rightarrow $ target states. They refer to identical eigen-energies
of internal states and identical translational states of $D$,
consistent with the indistinguishability requirement discussed
above. To further ensure such quantum interference, the coefficients
associated with these nondiagonal terms, namely,
$\Smatrix(E_{C},n_{C};E_{D},n_{D}|\Omega)
\Smatrix^{*}(E_{C},n_{C}';E_{D},n_{D}|\Omega)$ ($n_{C}\neq n_{C}')$,
should be nonzero. Since $n_{C}$ and $n_{D}$ usually refer to
quantities such as the parity or the projection of the total angular
momentum onto a space-fixed axis, at first glance it may appear that
$\Smatrix(E_{C},n_{C};E_{D},n_{D}|\Omega)
\Smatrix^{*}(E_{C},n_{C}';E_{D},n_{D}|\Omega)=0$  for
 $n_{C}\neq n_{C}'$, because
either $n_{C}n_{D}$ or $n_{C}+n_{D}$ should be conserved in the
scattering processes. But this is not true. Instead, we note (1)
that  the conservation laws of parity and angular momentum  only require
 \begin{eqnarray}
 \int d\Omega\
 \Smatrix(E_{C},n_{C};E_{D},n_{D}|\Omega)
 \Smatrix^{*}(E_{C},n_{C}';E_{D},n_{D}|\Omega)=0, \ n_{C}\neq n_{C}',
 \end{eqnarray}
 and (2) that for any particular
direction characterized by $\Omega$ [e.g., the direction strongly
preferred by the second collision via the $\Omega$ dependence of
$T^{n_{C}, n_{C}'}_{E_{C}, E_{D}, n_{D}}(\Omega)$],
$\Smatrix(E_{C},n_{C};E_{D},n_{D}|\Omega)
\Smatrix^{*}(E_{C},n_{C}';E_{D},n_{D}|\Omega)$ can be nonzero.
Hence, bimolecular collisions can create a certain type of
potentially useful quantum coherence between degenerate molecular
states, but in a very subtle manner. This result may be of interest
to considerations on controlling bimolecular reactions
\cite{paulPRL}, where superposition states consisting of degenerate
internal states are used. Nevertheless, in general, quantum
coherence between degenerate states is of limited use.

We summarize the results to conclude this section.
We have considered the collision of $A+B\rightarrow C+D$ at fixed
total energy and total momentum, and asked if the coherence
created in the products of this collision can be used in a subsequent
collisional step to do coherent control. To be experimentally feasible
this would mean that we would be trying to use
the coherence established in {\it one} of the product molecules in the subsequent
step. We would call this "useful coherence". Our results show that
(a) bimolecular collisions, such as multi-channel collision
$A + B\rightarrow C + D$,
cannot produce useful molecular translational coherence,
(b) bimolecular collisions cannot
achieve useful coherence between  non-degenerate ro-vibrational states,
(c) bimolecular collisions
cannot produce useful coherence between different product states of
various translational and ro-vibrational states,
and (d) bimolecular collisions, however, can produce
certain useful coherence between degenerate molecular states, such as those
with different parities or projections of
the total angular momentum onto a space-fixed axis.

\section{Indistinguishability and Interference in Coherent Control of Photochemical Processes}
\label{s4-3}

In this section we consider a representative phase-sensitive
coherent control scheme, namely, a particular two-pulse coherent control
scenario \cite{seideman,tannor-rice}, to examine the fundamental issue of
indistinguishability and interference in the coherent control of
photochemical processes. We then carry out similar examinations of
two phase-insensitive interference control scenarios.

\subsection{Classical Treatment of Laser Fields in Two-pulse Coherent Control}

Consider a classical linearly polarized electric  field $E(t)$ incident
on an initially bound molecule. The
molecule is assumed to be in an eigenstate $|E_{0}\rangle$ of the
molecular Hamiltonian $H_{M}$.
The overall Hamiltonian, in the dipole approximation, is then given by
\begin{eqnarray}
H=H_{M}-\hat{d}[E(t)+E^{*}(t)],
\end{eqnarray}
where $\hat{d}$ is dipole moment operator along the electric field.
In a two-pulse control scenario, the external field consists of two
separated Gaussian pulses $E_{x}(t)$ and $E_{d}(t)$ centered at
$t=t_{x}$ and $t_{d}$, respectively. The Fourier transform of
$E_{x}(t)$ and $E_{d}(t)$ is given by $E_{x}(\omega)$ and
$E_{d}(\omega)$. The first pulse $E_{x}(t)$ induces a transition to
a superposition  of bound excited molecular states and the second
pulse dissociates the molecule by further exciting it to the
continuum. Both fields are chosen to be sufficiently weak to apply
first-order perturbation theory.

Assuming that the first pulse encompasses only two $|E_{1}\rangle$
and $|E_{2}\rangle$
excited states, the superposition state thus prepared is given by
\begin{eqnarray}
|\phi(t)\rangle=|E_{0}\rangle \exp(-iE_{0}t/\hbar)+c_{1}|E_{1}\rangle
\exp(-iE_{1}t/\hbar)+c_{2}|E_{2}\rangle
\exp(-iE_{2}t/\hbar),
\end{eqnarray}
where
\begin{eqnarray}
c_{m}=\frac{\sqrt{2\pi}}{i\hbar}d_{m,0} E_{x}(\omega_{E_{m}E_{0}}),
\ \ m=1,2,
\end{eqnarray}
with $\omega_{E_{m}E_{0}}\equiv(E_{m}-E_{0})/\hbar$, and $d_{m,0}\equiv\langle
E_{m}|\hat{d}|E_{0}\rangle$.
This superposition state is subjected to a second pulse after a
time delay $(t_d-t_x)$.
When the second-pulse is completed, the system wavefunction is given by
\begin{eqnarray}
|\psi(t)\rangle=|\phi(t)\rangle+\sum_{n,q}\int dE\ B(E,n,q|t)|E, n,
q^{-}\rangle \exp(-iEt/\hbar), \label{classical}
\end{eqnarray}
where $E$, $n$, and $q$ denote the eigenenergy, the quantum numbers other
than the energy, and
the arrangement index for the eigenfunction $|E,n,q^{-}\rangle$ in the continuum.
The probability of observing the state $|E,n,q^{-}\rangle$
in the remote future is given by
\begin{eqnarray}
P(E,n,q)&=&\lim_{t\rightarrow\infty}\langle\psi(t)|E,n,q^{-}\rangle\langle
E,n,q^{-}|\psi(t)\rangle\nonumber \\
&=&|B(E,n,q|t=\infty)|^{2} \nonumber \\
&=&\frac{2\pi}{\hbar^{2}}\left|
\sum_{m=1,2}c_{m}\langle E,n,q^{-}|\hat{d}|E_{m}\rangle
E_{d}(\omega_{EE_{m}})\right|^{2} \nonumber \\
&=& \frac{2\pi}{\hbar^{2}}\left\{|c_{1}|^{2}d^{q}_{1,1}\left|E_{d}(\omega_{EE_{1}})\right|^{2}+
|c_{2}|^{2}d^{q}_{2,2}\left|E_{d}(\omega_{EE_{2}})\right|^{2}\right\}
+ I_{12}(t_{d}-t_{x}),
\label{cterm}
\end{eqnarray}
with
\begin{eqnarray}
\langle E_{1}|\hat{d}|E_{0}\rangle\langle E_{0}|\hat{d}|E_{2}\rangle
\equiv|\langle
E_{1}|\hat{d}|E_{0}\rangle\langle E_{0}|\hat{d}|E_{2}\rangle|\exp(i\theta),
\end{eqnarray}
\begin{eqnarray}
|d^{q}_{i,m}(E)|\exp[i\alpha^{q}_{i,m}(E)] \equiv
\langle E,n,q^{-}|\hat{d}|E_{i}\rangle\langle E_{m}|\hat{d}|E,n,q^{-}\rangle,
\end{eqnarray}
and
\begin{eqnarray}
I_{12}(t_{d}-t_{x})& =& \frac{4\pi}{\hbar^{2}}\left|c_{1}c_{2}^{*}E_{d}({\omega_{EE_{1}}})E_{d}^{*}(\omega_{EE_{2}})\right||d^{q}_{1,2}| \nonumber \\
&&\times \cos\left[\omega_{E_{2}E_{1}}(t_{d}-t_{x})
 +\alpha^{q}_{1,2}(E)+\theta\right],
 \label{interm}
\end{eqnarray}
Clearly, by changing the time delay between the two pulses or the
ratio between $c_{1}$ and $c_{2}$, one can manipulate the
interference term $I_{12}(t_{d}-t_{x})$. Moreover,  due to the
presence of the molecular phase $\alpha^{q}_{1,2}$, the interference
may be constructive for one arrangement while being destructive for
other arrangements. Selectivity can thus be achieved through the
manipulation of quantum interference and excellent control has been
predicted \cite{seideman} and observed experimentally \cite{gerber}.

\subsection{Fully Quantized Theory of Two-pulse Coherent Control}
The above classical treatment of light fields has several advantages.
For example, it shows, in a very simple manner,
the source of the interference and
how it can be manipulated experimentally.   It can also be readily extended to
cases in which laser incoherence is present.  However,
as the apparatus for the double-slit experiment is
treated quantum mechanically in Bohr's defense of the consistency of
quantum mechanics \cite{jammer,zurek41}, it is advantageous here to
consider a fully quantized theory of the two-pulse coherent control scheme.
As will become evident,  by also quantizing light fields,
we can readily examine the implications of
molecule-photon entanglement, the
indistinguishability between independent excitation pathways, and hence
expose the key difference
between bimolecular collisions and laser-molecule interaction.

Consider then a molecule subjected to a quantized electromagnetic field.
The total Hamiltonian including the molecular Hamiltonian $H_{M}$, the
radiation Hamiltonian $H_{R}$, and the interaction Hamiltonian $H_{I}$,
in the
Schr\"{o}dinger picture,  is given by \cite{loudon}
\begin{eqnarray}
H&=&H_{M}+H_{R}+H_{I}\equiv H_{0}+H_{I} \nonumber \\
&=&\sum_{{\bf k}}\hbar\omega_{{\bf
k}}\akh\ak+\sum_{j}E_{j}|j\rangle\langle j|+i\sum_{{\bf
k}}\sum_{mn}\sqrt{\frac{\hbar\omega_{{\bf k}}}{2\epsilon_{0}V}}
\cdot d_{mn} (\ak-\akh)|m\rangle\langle n|, \label{hh}
\end{eqnarray}
where the electrical field is assumed to be in the same direction as
the molecular dipole moment, $d_{mn}\equiv\langle E_{m}|\hat{d}|E_{n}\rangle$,
$a_{{\bf k}}$ and $a_{{\bf k}}^{\dagger}$ are the photon
annihilation and creation
operators for the frequency component $\omega_{{\bf k}}$,
$V$ is the quantization volume for the quantum field, and $\epsilon_{0}$
is the permittivity of the vacuum.

Given the Hamiltonian in Eq. (\ref{hh}), first-order perturbation theory gives
\begin{eqnarray}
\exp(-iHt/\hbar)&=&\exp(-iH_{0}t/\hbar)\{1+\frac{1}{\hbar}\sum_{{\bf
k}}\sum_{mn}d_{mn}|m\rangle\langle n|\sqrt{\frac{\hbar\omega_{{\bf k}}}{2\epsilon_{0}V}}[\ak\frac{
\exp[i(\omega_{E_{m}E_{n}}-\omega_{{\bf k}})t]}{i
(\omega_{E_{m}E_{n}}-\omega_{{\bf k}}-i\epsilon)}\nonumber \\
&& -\ \akh
\frac{
\exp[i(\omega_{E_{m}E_{n}}+\omega_{{\bf k}})t]}{i
(\omega_{E_{m}E_{n}}+\omega_{{\bf k}}-i\epsilon)}]\},
\label{firstorder}
\end{eqnarray}
where $\epsilon$ finally goes to $0^{+}$.
We define the time $t=0$ as that after which the first pulse is over.
Further, in addition to the assumptions of the classical field treatment,
we assume that the quantum
state of the first pulse at $t=0$ would be given by $|\psi^{x}\rangle$
if there were no laser-molecule interaction.
Then, the wavefunction for the entire system at $t=0$ is given by
(in the rotating wave approximation)
\begin{eqnarray}
|\Psi(t=0)\rangle & = &|\psi^{x}\rangle\otimes
|E_{0}\rangle +
\frac{1}{\hbar}(d_{10}\hat{A}_{10}|\psi^{x}\rangle)\otimes
|E_{1}\rangle  \nonumber \\
&& +\ \frac{1}{\hbar}(d_{20}\hat{A}_{20}|\psi^{x}\rangle)\otimes
|E_{2}\rangle,
\end{eqnarray}
where operators $\hat{A}_{j0}$ are
\begin{eqnarray}
\hat{A}_{j0}\equiv\sum_{{\bf
k}}\sqrt{\frac{\hbar\omega_{{\bf k}}}{2\epsilon_{0}V}}\frac{
\ak}{i(\omega_{E_{j}E_{0}}-\omega_{{\bf
k}}-i\epsilon)}, \ j=1,2.
\end{eqnarray}
For the second laser pulse,
the quantum state of light is assumed to be
$|\psi^{d}\rangle$ at $t=0$.  Applying first-order perturbation theory
a second time gives rise to the wavefunction $|\Psi(t)\rangle$ for
the entire system at any time $t\ge 0$. For the sake of
comparison with the classical treatment, we also assume
that only the $|E_{1}\rangle$ and $|E_{2}\rangle$ levels
contribute to the photodissociation probabilities.
Then one finds
\begin{eqnarray}
 \lim_{t\rightarrow + \infty} |\Psi(t)\rangle& \rightarrow & \exp(-iH_{0}t/\hbar)|\Psi(0)\rangle
 \nonumber \\ & & -\exp(-iEt/\hbar)
\frac{1}{\hbar^{2}}\sum_{n,q}\int dE |E,n,q^{-}\rangle\nonumber \\
 & & \otimes\ [d_{E,1}^{n,q}d_{10}(\hat{B}_{E,1}|\psi^{d}\rangle)\otimes
(\hat{A_{10}}|\psi^{x}\rangle)  \nonumber \\
& &\ \ \ + \
d_{E,2}^{n,q}d_{20}(\hat{B}_{E,2}|\psi^{d}\rangle)\otimes
(\hat{A}_{20}|\psi^{x}\rangle)],
\label{quan}
\end{eqnarray}
where $d_{E,i}^{n,q}\equiv \langle E,n,q^{-}|\hat{d}|E_{i}\rangle$, and
where $\hat{B}_{E,j}$ is given by
\begin{eqnarray}
\hat{B}_{E,j}\equiv \sum_{{\bf k}}\sqrt{\frac{\hbar\omega_{{\bf
k}}}{2\epsilon_{0}V}}\frac{\ak}{i(
\omega_{EE_{j}}-\omega_{{\bf k}}+i\epsilon)}, \ j=1,2.
\end{eqnarray}

\subsection{Interference and Indistinguishability in Two-pulse Coherent Control}
Equation (\ref{quan}) shows
that post-laser excitation the molecular state is
entangled with both
the first and second light fields. It is therefore possible that with
this entanglement
one can identify the excitation pathway trhough which the molecule is
dissociated, i.e., by either
the route $|E_{0}\rangle \rightarrow |E_{1}\rangle \rightarrow |E\rangle$
or the route $|E_{0}\rangle \rightarrow
|E_{2}\rangle \rightarrow |E\rangle$. This could result in the loss of
interference.
However, unlike the bimolecular collision case analyzed in Sec. II,
here the photon states that are
entangled with the molecular states are usually not orthogonal. This suggests that
the two independent excitation pathways can still
have a high degree of indistinguishability. Specifically, if we define
\begin{eqnarray}
|\psi_{1}\rangle&\equiv&\frac{\exp(-iEt/\hbar)}{\hbar^{2}}
\int dE\sum_{n,q}|E,n,q^{-}\rangle \nonumber \\
& & \otimes\  d_{E,1}^{n,q}d_{10}(\hat{B}_{E,1}|\psi^{d}\rangle)\otimes
(\hat{A_{10}}|\psi^{x}\rangle), \nonumber \\
& & \nonumber  \\
|\psi_{2}\rangle&\equiv&\frac{\exp(-iEt/\hbar)}{\hbar^{2}}\int dE \sum_{n,q}|E,n,q^{-}\rangle \nonumber \\
&& \otimes\
d_{E,2}^{n,q}d_{20}(\hat{B}_{E,2}|\psi^{d}
\rangle)\otimes
(\hat{A_{20}}|\psi^{x}\rangle),
\label{twopsis}
\end{eqnarray}
then the continuum part of $|\Psi(t)\rangle$ is just a superposition
of these two states, corresponding to
different contributions from independent excitation pathways.
The interference of these two states for the measurement of
the projector $|E,n,q^{-}\rangle\langle E,n,q^{-}|$ is given by
\begin{eqnarray}
& & \langle\psi_{1}|E,n,q^{-}\rangle\langle E,n,q^{-}|\psi_{2}\rangle+
\langle \psi_{2}|E,n,q^{-}\rangle\langle E,n,q^{-}|\psi_{1}\rangle
\nonumber \\
& =& \frac{1}{\hbar^{4}}\sum_{n}(d_{E,2}^{n,q})^{*}d_{E,1}^{n,q}d_{20}^{*}d_{10}\langle\psi^{x}|
\hat{A}_{2,0}^{\dagger}
\hat{A}_{1,0}|\psi^{x}\rangle\langle\psi^{d}|\hat{B}_{E,2}^{\dagger}
\hat{B}_{E,1}|\psi^{d}\rangle
+\ c.c.\ .
\label{qterm}
\end{eqnarray}

The first observation to be made from Eq. (\ref{qterm}) is that
an exact correspondence between classical and quantum treatment of the
light fields can be made under certain conditions. Suppose both
$|\psi^{x}\rangle$ and $ |\psi^{d}\rangle$ are products of coherent
states of light for different frequencies, i.e.,
\begin{eqnarray}
\hat{A}_{j0}|\psi^{x}\rangle&=&\sum_{{\bf
k}}\sqrt{\frac{\hbar\omega_{{\bf k}}}{2\epsilon_{0}V}}\frac{
\alpha_{{\bf k}}}{i(\omega_{E_{j}E_{0}}-\omega_{{\bf
k}}-i\epsilon)} |\psi^{x}\rangle \nonumber \\
&\equiv & \sqrt{2\pi}E_{x}^{q}(\omega_{E_{j}E_{0}})|\psi^{x}\rangle,
\nonumber \\
\hat{B}_{E,j}|\psi^{d}\rangle&=&\sum_{{\bf k}}\sqrt{\frac{\hbar\omega_{{\bf
k}}}{2\epsilon_{0}V}}\frac{ \beta_{{\bf k}}}{i(
\omega_{EE_{j}}-\omega_{{\bf k}}+i\epsilon)}
|\psi^{d}\rangle \nonumber \\ &\equiv& \sqrt{2\pi}E_{d}^{q}(\omega_{EE_{j
}})|\psi^{d}\rangle,
\label{qc}
\end{eqnarray}
where $\alpha_{{\bf k}}$ and $\beta_{{\bf k}}$ are the eigenvalues
of $\ak$ for the first and second light pulses, characterizing the
coherent states of light. One can then establish the equivalence
between Eqs. (\ref{qterm}) and (\ref{interm}) by requiring
$E_{x}^{q}(\omega_{E_{i}E_{0}})$ and $E_{d}^{q}(\omega_{EE_{i}})$
defined in Eq. (\ref{qc}) to be the same as the Fourier components
$E_{x}(\omega_{E_{i}E_{0}})$ and $E_{d}(\omega_{EE_{i}})$ of the
classical light fields. Of even greater interest is the implication
of this correspondence condition for the degree of
indistinguishability between the two independent excitation pathways
($|E_{0}\rangle \rightarrow |E_{1}\rangle\rightarrow E$  and
$|E_{0}\rangle\rightarrow
|E_{2}\rangle\rightarrow E$). Substituting Eq. (\ref{qc}) into Eq.
(\ref{twopsis}), one sees that $|\psi_{1}\rangle$ is absolutely
indistinguishable from $|\psi_{2}\rangle$ except for a c-number
phase factor, i.e., the degree of indistinguishability is one. Thus,
subject to the condition of Eq. (\ref{qc}), it is absolutely
impossible to tell which excitation pathway the molecule takes, even
after making precise measurements of the light fields. In other
words, the quantum-classical correspondence condition of Eq.
(\ref{qc})  here corresponds to the case of maximal degree of
indistinguishability between the two excitation pathways
$|E_{0}\rangle\rightarrow |E_{1}\rangle\rightarrow |E\rangle $
and $|E_{0}\rangle\rightarrow
|E_{2}\rangle\rightarrow E$. Hence, {\it coherent states of light for both
the preparation and dissociation pulses can first create and then
manipulate molecular coherence without leaking out any which-way
information}.

It then follows that quantum states of light in general may
provide some which-way
information in a molecular process that allows for multiple
excitation pathways. We find that this is evidently true in
some limiting cases.
Consider first a case in which the wavefunction $|\psi^{x}\rangle$ of
the first light field is an eigenstate $|n_{{\bf k}}\rangle $
of the photon number operator
$\akh\ak$ for $\omega_{{\bf k}}\approx \omega_{E_{1}E_{0}}$ and
still a coherent
state for other frequency components. Then, one can easily distinguish between the
two excitation possibilities  $|E_{0}\rangle\rightarrow
|E_{1}\rangle$ and $|E_{0}\rangle\rightarrow |E_{2}\rangle$,
by carring out a measurement of
the change in the number of photons with frequency
$\omega_{E_{1}E_{0}}$. That is,
if the number of photons with frequency
$ \omega_{E_{1}E_{0}}$ decreases by one, then the molecule must have been  excited to
$|E_{1}\rangle$, otherwise it must have been excited to $|E_{2}\rangle$.
According to the indistinguishability requirement for interference, this zero
indistinguishability completely destroys the interference.
Indeed, if we choose a complete set of distinguishing
projectors as $P_{{\bf k}}=|n_{{\bf k}}\rangle\langle n_{{\bf k}}|$, then
the corresponding degree of indistinguishability [see Eq. (\ref{defineu})]
between $|\psi_{1}\rangle$ and $|\psi_{2}\rangle$ in
Eq. (\ref{twopsis}) is zero;
and since $\langle n_{{\bf k}}|\hat{A}_{20}^{\dagger}\hat{A}_{10}
|n_{{\bf k}}\rangle=0$, the interference power for the projectors
$|E,n,q^{-}\rangle\langle E,n,q^{-}|$ is also zero.
Similarly, for the dissociation pulse, if $|\psi^{d}\rangle$
is  a  photon number eigenstate for one frequency component e.g.
$\omega_{EE_{1}}$,  interference will not exist
because we can  distinguish between $|E_{1}\rangle\rightarrow |E\rangle$ and
$|E_{2}\rangle\rightarrow |E\rangle$ by measuring the photon
number in the second light field.

One could argue that this example could be
intuitively understood in terms of the
photon number and photon phase
uncertainty relation say, $\delta N\delta \phi\approx 1 $. That is, a photon number eigenstate
($\delta N=0$)
gives the largest phase uncertainty ($\delta \phi \approx 2\pi$),  and
a large phase uncertainty destroys phase control.
However, it should be stressed that the
physics here is in fact more fundamental than is manifest in
this over-simplified  perspective. For example, consider
the two well-known quantum states of light, namely,
the even coherent states (ECS)
$|\psi^{ECS}\rangle=[2(1+\exp(-2\alpha^{2}))]^{-1/2}(|\alpha\rangle+
|-\alpha\rangle)$ and the  odd coherent states (OCS) $
|\psi^{OCS}\rangle=[2(1-\exp(-2\alpha^{2}))]^{-1/2}(|\alpha\rangle-
|-\alpha\rangle)$, where $\hat{a}|\alpha\rangle=\alpha|\alpha\rangle $ and $
\hat{a}|-\alpha\rangle=-\alpha|-\alpha\rangle$ \cite{optics}.   The photon
number distribution $P_{n}^{ECS}$ for ECS is given by $P_{n}^{ECS}$
$=[2\alpha^{2n}\exp(-\alpha^{2})]/[n!(1+\exp(-2\alpha^{2}))]$
if $n$ is even, and
$P_{n}^{ECS}=0$ if $n$ is odd; whereas the photon number distribution
 $P_{n}^{OCS}$ for OCS  is given by $P_{n}^{OCS}$=
$[2\alpha^{2n}\exp(-\alpha^{2})]/[n!(1-\exp(-2\alpha^{2}))]$
if $n$ is odd and $P_{n}^{OCS}=0$ if $n$ is even.
Note that such OCS and ECS can typically have $\delta N >>1$,
suggesting that the phase uncertainty could be very small ($\delta\phi<<1$)
if the uncertainty relation $\delta N\delta \phi\approx 1$ is applied.
However, here
loss of one photon in OCS (ECS) leads to a dramatic change
in the quantum states of light,  i.e.,
from only allowing for odd (even) numbers of photons to
only allowing for even (odd) numbers of photons.
Thus, if the preparation or dissociation pulse is given
by an OCS (ECS) for one frequency component (say,
$\omega_{E_{2}E_{0}}$), one can, in principle,
tell whether or not the molecule
has absorbed a photon of a certain frequency,
by a post-interaction measurement of
the even/odd property of the photon number distribution.  This implies
that in these cases  there should not be quantum interference, as a result
of complete distinguishability of multiple excitation pathways.
Indeed, based on the fact that
\begin{eqnarray}
 \langle\psi^{ECS}|\hat{a}|\psi^{ECS}\rangle=
\langle\psi^{OCS}|\hat{a}|\psi^{OCS}\rangle=0,
 \label{eocs}
\end{eqnarray}
one clearly sees that quantum interference given by Eq. (\ref{qterm})
should vanish if either the preparation or the dissociation pulse
is  given by OCS or ECS.

These results are also relevant to quantum computation.
Since quantum computation relies on coherently controlled evolution of atomic and molecular systems, the analysis
here suggests that nonclassical light fields may affect the reliability of a quantum computer by leaking out
some which-way information.  This is consistent with a recent study suggesting
that the quantum nature of light may have important implications for the limits of quantum computation \cite{QC}.

Comparing laser-molecule interaction considered here and bimolecular collisions
analyzed in Sec. II, we see that their key difference in quantum
coherence generation arises from two sources. First, a pulsed laser
field itself already carries coherence between different frequency
components and such coherence can be directly transferred to create
coherence between different molecular eigenstates. Second, a laser
field (e.g., when the quantum states of light are close to coherent
states) is somewhat of a classical object, thus the interaction between
a molecule and a laser field is more or less analogous to the
scattering between a molecule and a macroscopic object (e.g., a
classical diffraction slit in front of a double-slit plate). This is
in contrast to the scattering between two molecules (sub-microscopic
objects) assumed to be in momentum and energy eigenstates, where
quantum entanglement and hence the which-way information can be
easily established.

\subsection{Indistinguishability and Interference in Incoherent Control}
The physical picture established in the previous subsection applies
also to other weak-field coherent control scenarios [e.g., ``$1$
photon $+$  $3$ photons'' control, ``$1$ photon $+$  $2$ photons''
control, ``$ \omega_{1}+\omega_{2}$'' vs. ``$
\omega_{3}+\omega_{4}$'' control ($\omega_{1(2)}\neq\omega_{3(4)}$)]
with minor changes \cite{kral}. However, it remains to examine the
issue of indistinguishability and interference in several cases of
``incoherent interference control" schemes, where interference
between multiple pathways was found to be insensitive to laser
phases \cite{zchen,elliott}.

Consider first
the so-called ``$\omega_{1}+\omega_{2}$'' vs. `` $\omega_{2}+\omega_{1}$''
control \cite{elliott}.  In this case,
the first pathway starts with the excitation from state $|E_{0}\rangle$
to an intermediate state $|E_{1}\rangle$ by absorbing
a photon of frequency $\omega_{1}$, followed by the excitation from
state $|E_{1}\rangle$ to the target state $|E,n,q^{-}\rangle$ by a second
photon of frequency $\omega_{2}$. The second pathway proceeds through another
intermediate state $|E_{2}\rangle$ by first absorbing a photon
of frequency $\omega_{2}$,  and then being excited from $|E_{2}\rangle$ to $|E,n,q^{-}\rangle$
by absorbing a second photon
of frequency $\omega_{1}$.  Rigorously describing such laser-molecule
interaction requires a general resonant two-photon photodissociation
theory \cite{zchen2},  by which both level shifts and level widths
can be explicitly taken into account.  Nevertheless,  for the purpose here
it suffices to apply second-order perturbation theory
with the fully quantized Hamiltonian (\ref{hh}). Substituting (\ref{hh}) into the
following perturbation series
\begin{eqnarray}
 \exp(-iHt/\hbar)&=&\exp(-iH_{0}t/\hbar)[1-\frac{i}{\hbar}
\int^{t}_{0} dt_1\ \exp(iH_{0}t_{1}/\hbar)H_{I}\exp(-\epsilon t_{1})
\exp(-iH_{0}t_{1}/\hbar)\nonumber \\
& & -\frac{1}{\hbar^{2}}\int^{t}_{0}dt_{1}\int^{t_{1}}_{0}dt_{2}\
\exp(iH_{0}t_{1}/\hbar)H_{I}\exp(-\epsilon t_{1})
\exp(-iH_{0}t_{1}/\hbar) \nonumber \\
& & \times \exp(iH_{0}t_{2}/\hbar)H_{I}\exp(-\epsilon t_{2})
 \exp(-iH_{0}t_{2}/\hbar)],
\label{secondorder}
\end{eqnarray}
neglecting the first-order term (i.e., assuming that this term does
not contribute to photodissociation), and applying the rotating wave
approximation, one obtains the time-evolving state $|\Psi(t)\rangle$
for the entire molecule-field system. In particular, if initially
the state $|\Psi_{0}\rangle$ is a direct product state of the matter
wavefunction $|E_{0}\rangle$ and the light field wavefunction
$|\psi^{l}\rangle$, we have
\begin{eqnarray}
  \lim_{t \rightarrow +\infty}| \Psi(t)\rangle
&= &\exp(-iH_{0}t/\hbar)[|\Psi_{0}\rangle
 -\frac{1}{2\epsilon_{0} V\hbar}\sum_{{\bf k}{\bf k}'}
\sum_{j}\sum_{n,q}\int dE |E, n, q^{-}\rangle  \nonumber \\
&& \otimes\
\frac{\sqrt{\omega_{{\bf k}}\omega_{{\bf k}'}}
d_{Ej}^{nq}d_{j0}\hat{a}_{{\bf k}'}\hat{a}_{{\bf k}}}{(\omega_{EE_{0}}-\omega_{{\bf k}}-\omega_{{\bf k}'}+2i\epsilon)
(\omega_{EE_{j}}-\omega_{{\bf k}'}+i\epsilon)}|\psi^{l}\rangle],
\label{ex22}
\end{eqnarray}
where $\epsilon$ finally goes to $0^{+}$, the intermediate states are assumed to be $|E_{j}\rangle$,
and $d_{Ej}^{nq}$ and $d_{j0}$ are the associated transition dipole moments between
the intermediate state and initial or final states. As in the two-pulse control
case,
Eq. (\ref{ex22}) indicates that in general the final state is an entangled state
between the molecule and the light fields. The molecule-photon entanglement can, in general,
decrease the indistinguishability of the multiple pathways
associated with different intermediate states.  Interestingly,
this is not the case in the special situation used in the
``$\omega_{1}+\omega_{2}$'' vs. ``$\omega_{2}+\omega_{1}$''
scheme.  In this special case
there are only two near-resonant and dominant intermediate
bound states $|E_{1}\rangle$ and $|E_{2}\rangle$ of energy $E_{1}$ and $E_{2}$,
satisfying $\omega_{EE_{1}}=\omega_{E_{2}E_{0}}$. Hence,
\begin{eqnarray}
(\omega_{EE_{1}}-\omega_{{\bf
k}'}+i\epsilon)&=&(\omega_{E_{2}E_{0}}-\omega_{{\bf
k}'}+i\epsilon) \nonumber \\
& = & - (\omega_{EE_{2}}-\omega_{{\bf k}} +i\epsilon),
\label{relations}
\end{eqnarray}
where in obtaining the second equality we used
$(\omega_{EE_{0}}-\omega_{{\bf k}}-\omega_{{\bf
k}'}+2i\epsilon)\approx 0$ due to conservation of total energy.
Using Eq. (\ref{relations}) and manipulating the order of the sum in
Eq. (\ref{ex22}), the $E,n,q$-component of the wavefunction is found
to be
\begin{eqnarray}
   \lim_{t\rightarrow +\infty}|\Psi(t)\rangle_{E,n,q}&  =
   &\frac{\exp(-iEt/\hbar)}{2\epsilon_{0}\hbar V}
\sum_{{\bf k}{\bf
k}'}[
 \frac{-d_{E1}^{nq}d_{10}\sqrt{\omega_{{\bf k}}
\omega_{{\bf k}'}}|E,n,q^{-}\rangle\otimes \hat{a}_{{\bf k}'}\hat{a}_{{\bf
k}}|\psi^{l}\rangle}{(\omega_{EE_{0}}-\omega_{{\bf k}}-\omega_{{\bf k}'}+2i\epsilon)
(\omega_{EE_{1}}-\omega_{{\bf k}'}+i\epsilon)} \nonumber \\
& & +\
\frac{d_{E2}^{nq}d_{20}\sqrt{\omega_{{\bf k}}\omega_{{\bf k}'}}|E,n,q^{-}\rangle\otimes
\hat{a}_{{\bf k}'}\hat{a}_{{\bf
k}}|\psi^{l}\rangle}{(\omega_{EE_{0}}-\omega_{{\bf k}}-\omega_{{\bf k}'}+2i\epsilon)
(\omega_{EE_{1}}-\omega_{{\bf k}'}+i\epsilon)}].
\label{w1w2result}
\end{eqnarray}
Clearly, the first term on the right hand side of Eq.
(\ref{w1w2result}) represents the contribution from the first path
through the intermediate state $|E_{1}\rangle$, and the second term
represents the contribution from the second path associated with
$|E_{2}\rangle$. Remarkably, without any restriction on the form of
$|\psi^{l}\rangle$, the two terms in Eq. (\ref{w1w2result}) arising
from two excitation pathways are seen to be identical except the
c-number coefficients, i.e., the degree of indistinguishability of
these two components is one and the final molecular state
$|E,n,q^-\rangle$ is disentangled from the light field.   As such,
in the peculiar ``$ \omega_{1}+\omega_{2}$'' vs. ``$
\omega_{2}+\omega_{1}$'' control scheme, the final states arising
from the two independent pathways happen to be indistinguishable,
even after considering any possible molecule-field entanglement.
This is in sharp contrast to the two-pulse control case in which
only coherent states of light can guarantee the maximal degree of
indistinguishability.

Let us now consider a second incoherent interference control
example, here in the strong field \cite{zchen}.
In this case, the first pathway is simply a
direct excitation from an initial state to a target state,  and the
second pathway begins with an excitation from the initial state to
the same target state, followed by back and forth transitions
(induced by a strong field) between the target state and a third
intermediate state. Evidently,  classifying independent
pathways in this way is just a convenient zero-field picture for
understanding the associated quantum effects. These ``back and
forth'' transitions are simply fictitious excitation pathways in
a perturbation theory interpretation of
the excitation from the initial state to a dressed
target state. Given that it is absolutely impossible to distinguish
between fictitious pathways, even in principle, the maximal degree
of indistinguishability is automatically guaranteed, as is the
associated quantum interference.

To conclude, incoherent interference control is markedly
different from traditional coherent control, in that the former
utilizes a specific kind of quantum interference that results from
the absolute indistinguishability of multiple  excitation
pathways to the same target state.  It can therefore be anticipated that
incoherent interference control schemes will be applicable in cases involving
highly quantum states of light.

\section{Coherence Creation in Photodissociation Processes}
\label{s4-4}


In this section we apply insights from Sec. \ref{s4-3} and Sec. \ref{s4-4}  to
the related problem of quantum coherence creation in
photodissociation processes. On one hand, this problem is similar to
quantum coherence creation in bimolecular collisions
since usually there are two or more
products separating from one another in  the photodissociation processes.
On the other hand, the physics here involves the creation of molecular
coherence using light fields.

Suppose the photodissociation process is $AB\rightarrow
(AB)^{*}\rightarrow$ $C$ + $D$, where $(AB)^{*}$ represents, after
$AB$ absorbs one photon, the excited complex before it breaks apart
to form molecules $C$ + $D$. Conventional photodissociation
experiments that employ very long monochromatic laser pulses are
such that, essentially,  $(AB)^{*}$ has both definite energy and
momentum. Hence the process $(AB)^{*}\rightarrow C + D$ is precisely
the same as the second half of a bimolecular collision discussed in
Sec. \ref{s4-2}. It then follows that each photodissociation
fragment in this process cannot have  useful coherence between
different translational states, or between nondegenerate
rovibrational states \cite{kurizki}. This is fundamentally because
one can, in principle, obtain which-state information about one
fragment by measuring the other fragment, since the total momentum
and total energy are known. One can then predict that, although it
is common to have a broad rovibrational state distribution in a
photodissociation fragment, no definite phase relationships between
the non-degenerate rovibrational states of one individual product
molecule should be expected in traditional photodissociation
processes (i.e. that do not incorporate coincidence measurements).
This should be the case even when classical correlations in the
final state distributions between two fragments are weak.

As in bimolecular collisions, however, conventional photodissociation
with monochromatic sources
can still  generate coherence between {\it degenerate} internal
states of photodissociation fragments. For example,  consider a
laser field linearly polarized along the $x$ direction and examine the
coherence between states of $C$ with different $m_{z}$ (the quantum
number associated with the projection of the angular momentum onto
the $z$-axis). The associated selection rule is
$[m_{z}(C) + m_{z}(D)-m_z(AB) ]=+ 1$ or $[m_{z}(C) +
m_{z}(D)-m_z(AB)]=-1$. Apparently then,  knowledge of $m_{z}(D)$
cannot tell us the  precise value of $m_{z}(C)$ since there are still two
different possibilities.
Thus, a superposition state of $C$ comprising
two $m_{z}(C)$ components  can be created in such a
photodissociation process.

By contrast to the bimolecular analogy and monochromatic light
sources, femtosecond laser technology opens new interference
possibilities in photodissociation dynamics.   Ultrafast laser
pulses directly excite the molecule $AB$ into a coherent
superposition state of many ro-vibrational states embedded in the
continuum,  on a time scale much less than that of the
ro-vibrational motion. These photodissociation processes are no
longer analogous to the second half of bimolecular collisions due to
the large energy uncertainty introduced by ultrafast laser pulses.
Indeed, these processes are similar to those in various coherent
control scenarios where coherence is transferred from light fields
to molecules. As such, quantum states of photodissociation fragments
$C$ and $D$ can be largely free from entanglement in energy, i.e.,
by accurately measuring the energy of one fragment we can  not
necessarily specify the energy of the other fragment. Hence,
indistinguishability conditions can be satisfied, and ro-vibrational
coherence effects of one individual photodissociation fragment may
be created and further used for a second molecular process. One can
conclude that it is exactly the lack of entanglement between
photodissociation fragments that makes possible the previous
observation of coherent vibrational motion of photodissociation
fragments \cite{zewail,ruhman,ShapiroJPCA}. Note however, that under
the assumption that we can neglect the momentum carried by photons,
useful coherence in the translational motion of a product molecule
still cannot be created due to the momentum entanglement between
photodissociation fragments.

\section{Concluding Remarks}
\label{s4-5}

In this work, we have examined, within
in the context of coherent  control of molecular processes,
the subtle issue of indistinguishability and
interference between independent pathways.
Interference occurs only when independent pathways are indistinguishable.
Due to this indistinguishability requirement,
creating a useful
superposition state of nondegenerate molecular states or of momentum eigenstates
for subsequent coherent control
cannot be achieved by collisions between atoms or molecules (initially prepared
in energy and momentum eigenstates).
Coherence can, however,
be conveniently transferred from light fields to molecules.
This coherence transfer,
and the subsequent coherent control based on
this coherence transfer (as analyzed in the
two-pulse control as an example)
are best realized
by the most classical states of light, i.e.,  coherent states of light.

We have shown that quantum states
of light may suppress the extent of phase-sensitive coherent control
by leaking out some which-way information in quantum processes.
By contrast, incoherent interference control schemes are shown to
have automatically ensured the maximal degree of indistinguishability
between independent excitation pathways.  Thus, when extended
to a regime where the quantum nature of
light becomes important, some of the known optical control scenarios
should be effective whereas  some others may not work at all.

We have also discussed the implication of the relationship
between indistinguishability
and interference for understanding coherence creation
in photodissociation fragments.
It is shown that traditional photodissociation
processes with long monochromatic laser pulses cannot
create
useful molecular coherence between non-degenerate ro-vibrational
states due to the quantum entanglement between photodissociation fragments.
New possibilities afforded by femto-second
photodissociation processes are understood in terms of disentanglement in energy between
photodissociation fragments.

The essence of coherent laser control of atomic and molecular
processes can be often understood in parallel with a double-slit
quantum interference experiment. This work further strengthens this
analogy. In particular, it now becomes clear that laser fields in
coherent phase control play the similar role as the classical
diffraction single-slit plate and the classical double-slit plate,
in that laser fields first create independent pathways and then
recombine these pathways to generate quantum interference. Though
somewhat counterintuitive, we can now conclude that, due to the
unavoidable quantum entanglement between quantum systems that have
interacted in the past, the successful creation of quantum coherence
or interference phenomenon in a quantum system often involves a
macroscopic object that is describable by classical physics, such as
classical slits in the case of double-slit matter-wave interference
experiments, or a sufficiently classical electromagnetic field in
the case of coherent interference control experiments. As is now
clear, roughly speaking this is because a classical object can
remain disentangled with the quantum system of interest and
independent pathways are therefore created without leaking out
which-way information. Such a role of classical objects in
generating useful quantum coherence, we believe, deserves more
attention in understanding the connection between classical and
quantum worlds.

\section{Acknowledgments}
This work is largely based on one chapter of the first author's
Ph.D thesis at the University of Toronto \cite{Gongthesis}.  This
work was supported by the U.S. Office of Naval Research and the
Natural Sciences and Engineering Research Council of Canada.

\pagebreak


\begin{thebibliography}{100}
\bibitem{kral} Note that the situation can be different if the goal is to create entangled
matter-photon states.  One may refer to an interesting article by P.
Kr\'{a}l, I. Thanopulos, and M. Shapiro, \pra{\bf 72}, 020303
(2006), where ``1 photon + 2 photon" control was considered and only
one of the two light fields is quantized.
\end{thebibliography}
\end{document}